\documentclass[aps,prd,reprint,superscriptaddress,nofootinbib,amsmath,amssymb,floatfix]{revtex4-2}
\usepackage{graphicx}
\usepackage{braket}
\usepackage{simpler-wick}
\usepackage[hidelinks,unicode]{hyperref}
\hypersetup{
  pdftitle={Quantum channel capacity of traversable wormholes},
  pdfauthor={Jingru Lu, Zhenbin Yang, and Jianming Zheng},
  pdfsubject={Phys. Rev. D 114, 106025 (2026), DOI: 10.1103/7trm-rykk}
}
\newcommand{\e}{\mathrm{e}}

\begin{document}

\title{Quantum channel capacity of traversable wormholes}
\author{Jingru Lu}
\affiliation{Institute for Advanced Study, Tsinghua University, Beijing 100084, China}
\author{Zhenbin Yang}
\affiliation{Institute for Advanced Study, Tsinghua University, Beijing 100084, China}
\affiliation{Peng Huanwu Center for Fundamental Theory, Hefei, Anhui 230026, China}
\author{Jianming Zheng}
\affiliation{Institute for Advanced Study, Tsinghua University, Beijing 100084, China}
\affiliation{Department of Physics and Astronomy, University of California, Davis, California 95616, USA}

\date[]{Published in \href{https://doi.org/10.1103/7trm-rykk}{Phys. Rev. D \textbf{114}, 106025} on 23 November 2026}

\begin{abstract}
We formulate the Gao-Jafferis-Wall traversable wormhole protocol as a quantum channel and compute its quantum channel capacity. We show that this capacity is governed by the time derivative of an out-of-time-ordered correlator, hence by operator size growth in the holographic dual, and that its growth is bounded above by the Einstein gravity limit. The channel capacity therefore provides a natural benchmark for quantum simulations of traversable wormholes.
\end{abstract}
\maketitle

\section{Introduction}

The traversability of spacelike wormholes has been a tantalizing riddle in quantum gravity. In classical general relativity, traversable wormholes are forbidden by the averaged null energy condition~\cite{Gao:2000ga}. However, by adding exotic matter quantum fields, which violate the averaged null energy condition, their existence is allowed. As a simple instance, one can consider coupling the two boundaries of an eternal black hole in anti-de Sitter spacetime by a double trace deformation, proposed by Gao \textit{et al.}~\cite{Gao:2016bin}.

In holography, this bulk picture of traversable wormholes has a dual description in the boundary quantum system. The quantum state of a double-sided black hole corresponds to a thermofield double (TFD) state in the dual quantum mechanical system~\cite{Maldacena:2001kr}. The TFD state lives in a doubled Hilbert space. In this sense, the two copies could be freely coupled. Alternatively, this coupling can be viewed as an information transfer process between the two boundary systems. With the help of the large number of entangled pairs in the TFD state, simple double-sided couplings can serve as measurements and unitary operations on the two sides, such that a standard quantum teleportation protocol is established~\cite{Maldacena:2017axo,Susskind:2017nto}. In the bulk, this quantum information-theoretic process has a geometric picture: the particles carrying the information simply travel along the Einstein-Rosen bridge between two asymptotic regions.

The traversable wormhole protocol has two key advantages over other existing teleportation protocols. Compared with ordinary quantum teleportation, the required unitary may act on any pair of qubits, not necessarily the pair that initially carries the information. This is possible because the protocol implements many-body teleportation through operator scrambling~\cite{Maldacena:2017axo,Gao:2019nyj}. Compared with the Kitaev-Yoshida protocol~\cite{Yoshida:2017non}, which uses a Grover-search-based construction in a general chaotic many-body system, the traversable wormhole protocol is one-shot: a single double trace deformation can transmit a large amount of information between the two sides. In this work, we quantify the maximal information transfer in this one-shot process using the quantum channel capacity, and show that it provides a quantitative benchmark for quantum simulations of traversable wormholes~\cite{Brown:2019hmk, Jafferis:2022crx,Kobrin:2023rzr}.

We will revisit the Gao-Jafferis-Wall setup in two-dimensional anti-de Sitter spacetime ($\mathrm{AdS}_2$)~\cite{Gao:2016bin,Maldacena:2017axo}. As initiated in~\cite{Bao:2018msr,Bao:2019rjy}, one can formulate the traversable wormhole as a quantum channel: The initial states to be transported are prepared by inserting various conformal primary operators on the right boundary of a two-sided black hole in $\mathrm{AdS}_2$, and the double trace deformation serves as the unitary operator on the bipartite Hilbert space. After tracing out the right black hole, this defines a standard quantum channel that maps from the initial states to the left black hole.

In this setup, we calculate the quantum channel capacity using coherent information of the quantum channel. Interestingly, we find an explicit relation between this quantity and the out-of-time-ordered correlation function (OTOC), which is a standard probe of quantum chaos. This demonstrates that the channel capacity is accounted for by the operator size growth of black holes~\cite{Roberts:2018mnp,Qi:2018bje}, in comparison with the operator size winding mechanism proposed in~\cite{Brown:2019hmk}. We comment on the effect of incorporating stringy corrections in the gravitational scattering, which leads to submaximal chaos and a slower growth of quantum channel capacity.

\textit{Connection to other works.} Reference~\cite{Schuster:2021uvg} discussed the channel capacity of the wormhole teleportation protocol from the perspective of general operator growth dynamics, with a particular focus on the scaling of the channel capacity with the number of coupled qubits ($K$). We worked in a different region where $K$ was taken to infinity and studied the time dependence of the quantum channel capacity.

\section{Traversable wormhole in \texorpdfstring{$\mathrm{AdS}_2$}{AdS2}}

\begin{figure}
    \centering
    \includegraphics[width=0.65\linewidth]{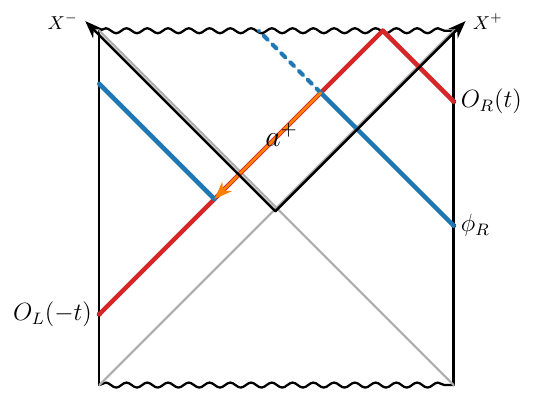}
    \caption{The Penrose diagram of traversable wormhole. We use orange line to denote $a^{+}$, which is the null shift experienced by $\phi$ particles due to the shockwave created by $O$ particles along $X^{+}$ direction. In the diagram, we ignore the backreaction of $\phi$ particles on $O$ particles. }
    \label{traversable}
\end{figure}

We now briefly review the Gao-Jafferis-Wall traversable wormhole in $\mathrm{AdS}_2$~\cite{Maldacena:2017axo}. We start with a thermofield double state with inverse temperature $\beta$:
\begin{equation}\label{TFD}
    |\text{TFD}\rangle=\frac{1}{\sqrt{Z}}\sum_n\e^{-\beta E_n/2}|E_n\rangle_L|\overline{E}_n\rangle_R\in\mathcal{H}_L\otimes\mathcal{H}_{R}.
\end{equation}
Here $E_n$ are the energy levels of the system. $\mathcal{H}_{L,R}$ are the Hilbert spaces of the left and right black holes and $Z$ is a normalization factor. The infalling particles are modeled by operators $\phi_R^a$ inserted at $t_R=0$ that carry a flavor index $a=1,\ldots,k$, with the same conformal dimension $\Delta_{\phi}$.
The double trace deformation can be modeled as the following unitary:
\begin{equation}\label{double trace}
    U=e^{i gV},\quad V(t)=\frac{1}{K}\sum_{i=1}^KO^i_L(-t)O^i_R(t),
\end{equation}
where $O^i_{L,R}$ are simple matter operators in the boundary quantum mechanics system (such as the Sachdev--Ye--Kitaev (SYK) model~\cite{Maldacena:2016hyu,Kitaev:2017awl}) with the same scaling dimension $\Delta_O$. For $t$ around scrambling time $t\sim \log\frac{1}{G_N}$, these fields backreact strongly on the geometry such that an infalling particle, which represents the information to be sent, experiences a null shift near the horizon toward the other side (see Fig.~\ref{traversable}). We take $K$ to infinity and $g$ to be large and finite, while keeping $k$ as a tunable parameter. In this limit, we can ignore particle creation from the double trace deformation, and treat $V$ geometrically as creating a shockwave~\cite{Maldacena:2017axo}.

The gravitational interaction between the shockwave and probe particles at large center-of-mass energy is described by the Dray--'t~Hooft $S$-matrix. The null shift created by $O_{L,R}$ acts on the in-state through the following unitary operator:
\begin{equation}
    S=e^{i G_N \hat{P}_+\hat{P}_-}.
\end{equation}
Here $\hat{P}_+$, $\hat{P}_-$ are the null momenta carried by the bulk quanta corresponding to $\phi$ and $O$. We work in the frame where $\phi$s are not boosted, and $\phi$ particles experience a null shift $a^+$ in the $x^+$ direction
\begin{equation}
    a^+\sim p_-G_N\sim -gG_Ne^t,\quad S=e^{i \hat{P}_+a^+}.
\end{equation}
Here $a^+$ is linear in $G_N$, meaning that the null shift is a gravitational backreaction effect.

The energy of the $\phi$ particle can backreact on the wormhole, which leads to an upper bound on the amount of information transferred between the two sides~\cite{Maldacena:2017axo}. One can roughly estimate the maximum number via the uncertainty principle. First, for a single particle, the traversal process can be modeled as a wave traveling through a window with width $a^+$. Therefore the momentum uncertainty of a single particle can be estimated as $\Delta p_+{a^+}\gtrsim 1$. We also require the backreaction from the $\phi$ particles to be less than order one, then the total momenta $|p_+|\lesssim \frac{1}{G_Ne^t}$\footnote{The upper index $-$ becomes $+$ in the lower one, thus the momentum related to $\phi$ particles moving along the $X^{-}$ direction is $p_{+}$.}. Based on these two inequalities, one can provide a rough bound on the information transfer, which is simply the number of transmitted $\phi$ particles
\begin{equation}
    \frac{|p_+|}{\Delta p_+}\lesssim \frac{|a_+|}{G_Ne^t}\sim g.
\end{equation}
In the remaining part of this paper, we would like to provide a more systematic study of the bound using quantum channel capacity, which is also generalizable to the case of nonmaximally chaotic systems, i.e., with stringy corrections.

\section{The quantum channel capacity of traversable wormhole}

We now introduce how the traversable wormhole can be formulated as a quantum channel and how its channel capacity is defined.

\begin{figure}[tbp]
    \centering
    \includegraphics[width=0.5\linewidth]{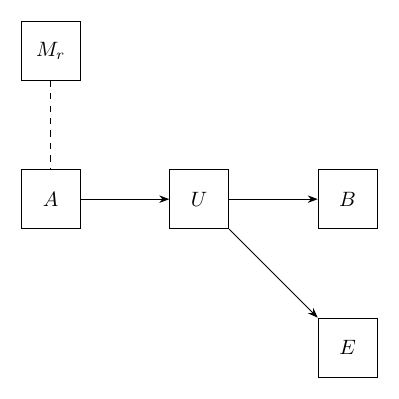}
    \caption{Schematic process of a quantum channel. $M_r$, $A$, $B$, $E$ denotes the Hilbert space $\mathcal{H}_{M_r}, \ \mathcal{H}_A, \ \mathcal{H}_B,\ \mathcal{H}_E$ respectively. And $\mathcal{H}_{M_r}$ is the reference system to purify the state $\rho_A\in\mathcal{H}_A$. $U$ is the isometric extension related to the quantum channel. $\mathcal{H}_E$ is the environment.}
    \label{fig:quantum_channel}
\end{figure}

Given any quantum channel $\mathcal{N}^{A\rightarrow B}:\mathcal{H}_A\rightarrow\mathcal{H}_B$, Stinespring dilation theorem~\cite{wilde2013quantum} states that there exists an isometric extension $U^{A\rightarrow BE}:\mathcal{H}_A\rightarrow\mathcal{H}_B\otimes\mathcal{H}_E$ such that $\operatorname{Tr}_E U^{A\rightarrow BE}(\rho_A)=\mathcal{N}^{A\rightarrow B}(\rho _A)$ for any $\rho_A \in \mathcal{H}_A$. In Fig.~\ref{fig:quantum_channel} we sketch the process of a quantum channel.
Using the reference system $\mathcal{H}_{M_r}$ and the isometric extension of the quantum channel, the process can be viewed as a map between pure states: $\ket{\psi}_{M_r A}\mapsto\ket{\phi}_{M_r BE}$. The quantum channel capacity quantifies the amount of information that we can transmit through the channel without entanglement assistance. Roughly speaking, the quantum channel capacity is the maximum number of qubits that can be transmitted with high fidelity per use of the quantum channel. The quantum channel capacity theorem~\cite{wilde2013quantum} states that the quantum capacity $C_Q(\mathcal{N})$ of a quantum channel is given by the regularized \textit{coherent information} of the channel
\begin{equation}
    C_Q(\mathcal{N})=Q_{\mathrm{reg}}(\mathcal{N}),
\end{equation}
where the regularized channel coherent information $Q_{\mathrm{reg}}(\mathcal{N})$ is defined as
\begin{equation}
    Q_{\mathrm{reg}}(\mathcal{N}^{A\rightarrow B})=\lim_{n\rightarrow\infty}\max_{A^n}\frac{1}{n}I_c(M_r^n \rangle B^n)_{\ket{\phi}_{M_r^nB^nE^n}},
\end{equation}
with $I_c(M_r  \rangle B)_{\ket{\phi}_{M_r BE}}$ being the coherent information with respect to state $\ket{\phi}_{M_r BE}$
\begin{equation}
    I_c(M_r \rangle B)_{\ket{\phi}_{M_r BE}}=S(B)-S(BM_r),
\end{equation}
with $S(X)$ denoting the von Neumann entropy of the system $X$. The regularized channel coherent information is defined by maximizing the coherent information over all initial states with many copies of the channel.
In general, it is very difficult to calculate the capacity using this formula directly. However, if the channel is additive, then the formula will be simplified
\begin{equation}
    C_Q(\mathcal{N}^{A\rightarrow B})=\max_{A} I_{c}(M_r \rangle B)_{\ket{\phi}_{M_r BE}},
\end{equation}
which means that one can consider one copy of the channel.

Now let us turn to our traversable wormhole case. We want to transfer information from the right-hand side of an AdS-Schwarzschild black hole to the left-hand side, so we consider the quantum channel as a map from a family of semiclassical bulk states with right-particle excitations to the left black hole. More precisely, we define $\mathcal{H}_A$ in Fig.~\ref{fig:quantum_channel} to consist of the family of semiclassical bulk states spanned by acting with $\phi^a_R$ on the right side of the black hole: $\lbrace|\phi^a_R\rangle=\frac{1}{\sqrt{Z_{\phi}}}\phi^a_R|\text{TFD}\rangle,\,a=1,\ldots,k\rbrace$,\footnote{$Z_{\phi}=\braket{\phi_R\phi_R}$ is the normalization factor, with $\bra{\text{TFD}}\phi_R^a\phi^b_R\ket{\text{TFD}}=\delta _{ab}\braket{\phi_R\phi_R}$.} and $\mathcal{H}_B$ to be the Hilbert space of the left black hole $\mathcal{H}_L$. The ``environment'' $\mathcal{H}_E$ is then given by the right black hole $\mathcal{H}_R$, such that the isometric extension of the quantum channel becomes the unitary $U$ in Eq.~\eqref{double trace}, with the time $t$ at which the double trace deformation acts viewed as a parameter of the quantum channel.\footnote{One could also enlarge the encoding space by including states prepared at different times. This then requires a maximization over time $t$ to obtain the channel capacity.} We can then introduce a reference system $\mathcal{H}_{M_r}$ with basis $\ket{\chi_a}_{M_r}$ and consider a general purified initial state $\ket{\alpha}=\sum_a \alpha^a |\phi^a_R\rangle |\chi_a\rangle_{M_r}$.

With this definition, the coherent information becomes the difference between the von Neumann entropy of the left and right black holes after the double trace deformation
\begin{equation}
    I_c(M_r \rangle L)=S(L)-S(R).
\end{equation}
Notice that we can do a common vacuum subtraction of $S(L)$ and $S(R)$ such that both entropies have a large $N$ limit. Notice also that the coherent information in this limit becomes the coherent information of bulk fields, although we will not use this in later calculation.

\begin{figure}
    \centering
    \includegraphics[width=0.5\linewidth]{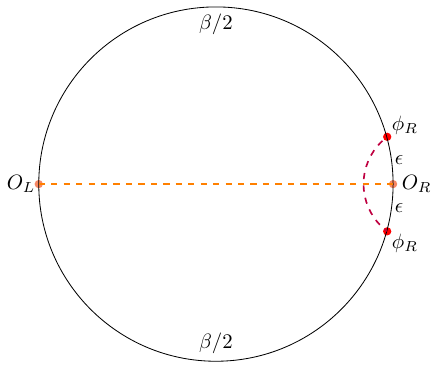}
    \caption{The OTOC configuration related to the coherent information, where $\epsilon $ is the Euclidean cutoff introduced for energy regularization.}
    \label{fig:OTOC}
\end{figure}

Later, we will argue that the traversable wormhole channel should be additive, as was argued previously in~\cite{Bao:2018msr}. Under this assumption, the quantum channel capacity is given by maximizing the coherent information discussed above over the initial state $|\alpha\rangle$. Appendix~\ref{app:replica} uses the replica trick to calculate both $S(L)$ and $S(R)$ in the large $N$ limit, and one finds
\begin{equation}
\begin{split}
        \Delta S(L)=\min(\Delta\langle  K_L\rangle,S(\alpha)+\Delta\langle  K_R\rangle);\\
        \Delta S(R)=\min(S(\alpha)+\Delta\langle  K_L\rangle,\Delta\langle  K_R\rangle),
\end{split}
\end{equation}
where $K_{L,R}$ is the modular Hamiltonian of the thermal state $\sigma$ of the original left/right black hole, and $\Delta\langle K_{L,R}\rangle=\braket{K_{L,R}}-\braket{K}_{\sigma}$ and $\Delta S(L/R)=S(L/R)-S(\sigma)$.
In particular, this means both $\Delta S(L)$ and $\Delta S(R)$ satisfy the Bekenstein bound $\Delta S\leq \Delta \langle K\rangle  $~\cite{Casini:2008cr}. $S(\alpha)$ is the entropy of the reference state $\rho_{\alpha}=\sum_a |\alpha^a|^2|\chi_a\rangle\langle\chi_a|$, and is bounded by $\log k$. $\Delta\langle K_{L,R}\rangle$ measures the energy of the $\phi$ particle in the left and right causal wedges; when $\Delta\langle K_{L}\rangle\gg\Delta\langle K_{R}\rangle$, the $\phi$ particle has been mostly transferred to the left side, the right system becomes thermal, namely $\Delta S(R)$ saturates to $\Delta \langle K_R\rangle$. Then the Bekenstein bound on $\Delta S(L)$ leads to an upper bound on the coherent information in terms of the boost energy $\beta B\equiv K_L-K_R$ of the final state, which can be saturated in the case of $\log k>\beta \langle B\rangle$, otherwise the coherent information is bounded by $\log k$. In the latter case, all information can be transferred. We will be interested in the former case. The same intuition also applies when we consider multiple copies of the channel, which implies $I_c(M_r^n \rangle L^n)\leq \beta n \braket{B}$. This in particular shows that the quantum channel is additive.  Thus, we have
\begin{equation}\label{saturate}
    \max_{\text{A}} I_c(M_r\rangle L)=\beta\braket{B}.
\end{equation}
The expectation value of the boost operator $B$ can be calculated straightforwardly to relate to the OTOC in the original TFD state
\begin{equation}
\begin{aligned}
    \braket{B}&=\bra{\phi} e^{-igV}B e^{igV}\ket{\phi}
    \\ &=\bra{\phi}e^{-igV}[B,e^{igV}]\ket{\phi}+\bra{\phi}B\ket{\phi}
    \\&= -g\partial_{t}\text{OTOC}(t)+\frac{1}{\beta}I_c^0.
\end{aligned}
\end{equation}
Here we use the property that $V$ acts geometrically as a shockwave and separate the action of the boost operator on the shockwave and the initial state respectively.
This property means there is only interaction between the $V$ and $\phi$, and in particular $\partial_t V$ commutes with $V$ in the correlation function, such that $e^{-igV}[B,e^{igV}]=-g\partial_t V$.
For a general large $N$, all-to-all interacting quantum mechanical systems such as the SYK model in different temperature regions, this property still holds, with $V$ and $\phi$ interacting through scramblon exchange~\cite{Kitaev:2017awl,Gu:2021xaj}, see Appendix~\ref{app:scramblon}.
This leads to the following OTOC correlator\footnote{Here we have dropped the flavor index.}
\begin{equation}\label{myOTOC}
    \text{OTOC}(t)\equiv\frac{\bra{\text{TFD}}\phi_R O_R(t)\phi_RO_L(-t)\ket{\text{TFD}}}{\langle\phi\phi\rangle}.
\end{equation}
The OTOC configuration is drawn in Fig.~\ref{fig:OTOC}.
The boost energy of the initial state is defined to be $I_c^0\equiv\beta \bra{\phi}B\ket{\phi}$.
The above discussion assumes $\langle B\rangle\gg 0,$ if $\langle B\rangle\ll 0$, then the information has not been transferred to the other side, so the channel capacity will be zero.
This leads to our main conclusion that the quantum channel capacity of the traversable wormhole protocol is given by the time derivative of the OTOC
\begin{equation}\label{eqn:QN}
    C_Q(\mathcal{N})=\max\lbrace -g \beta \partial_t\text{OTOC}(t)+ I^0_c,0\rbrace.
\end{equation}
We remind the reader that this is $C_Q(\mathcal{N})$ for the case of $\log k\gg \beta \langle B\rangle$. Otherwise $C_{Q}(\mathcal{N})=\log k$, namely all the information in the initial code subspace can be teleported. We remark again that this is a general result for large $N$, all-to-all interacting quantum mechanical systems.

It is interesting to examine the dependence of $C_Q(\mathcal{N})$ on $t$. The above formula shows that $C_Q(\mathcal{N})$ has an initial growth near the scrambling time of the form $g\lambda  G_N e^{\lambda t}$, controlled by the Lyapunov exponent $\lambda$. In particular, it is bounded by Einstein gravity due to the chaos bound $\lambda \leq \frac{2\pi}{\beta}$~\cite{Maldacena:2015waa}. In this sense, pure Einstein gravity, or a maximally chaotic system, leads to the fastest growth of the quantum channel capacity.

In Jackiw-Teitelboim (JT) gravity, the full analytic result of the channel capacity in the eikonal region can be expressed using the exact result of the OTOC in~\cite{Maldacena:2016upp}:\footnote{Here we have set $\beta=2\pi$ for convenience.}
\begin{figure}
    \centering
    \includegraphics[width=.49\columnwidth]{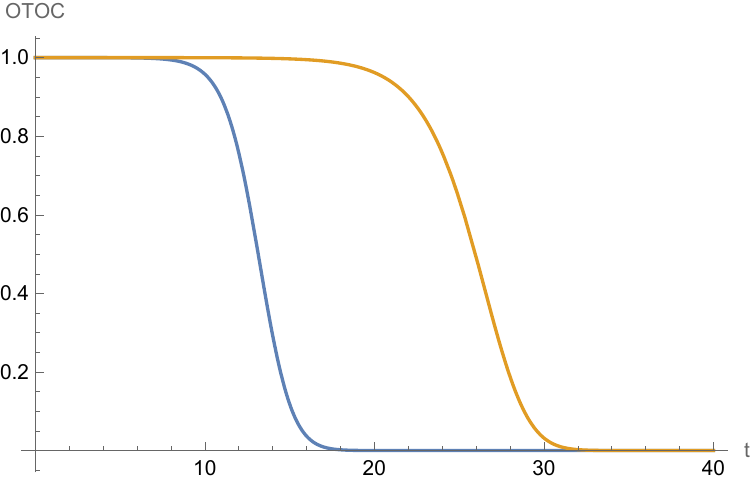}\hfill
    \includegraphics[width=.49\columnwidth]{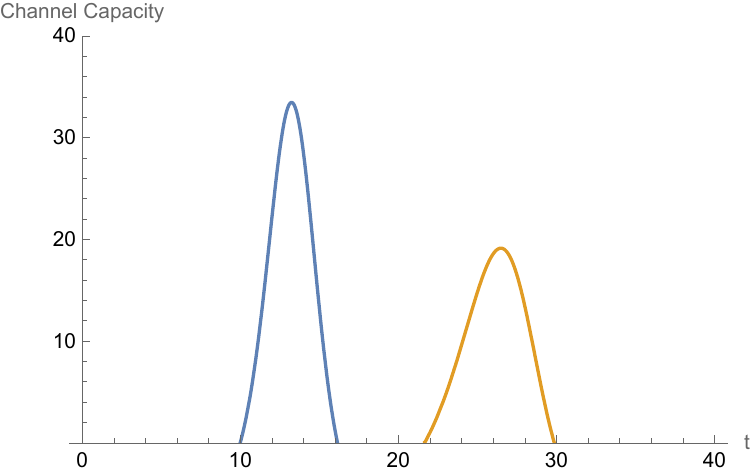}
    \caption{Left: plot of the relation between the OTOC and $t$ with maximal chaos and stringy corrections, respectively, where $g=100$, $\epsilon=1$, $\Delta_\phi=\Delta_O=1$ and $4C=10^6$. Right: plot of the relation between $C_Q(\mathcal{N})$ and $t$. The blue curves represent the OTOC and $C_Q(\mathcal{N})$ of JT gravity and the yellow curves represent the case with stringy correction ($a=1/2$).}
    \label{fig:S_L}
\end{figure}
\begin{equation}
\begin{aligned}
&C_Q(\mathcal{N}) = \max\biggl\{
  8\pi g\Delta_O\Delta_\phi\frac{2^{2\Delta_\phi-2\Delta_O}}{T^{2\Delta_\phi}} \\
  &\qquad {}\times U\!\left(2\Delta_\phi+1,2(\Delta_\phi-\Delta_O)+1,\frac{2}{T}\right) \\
  &\qquad {}-\frac{2\pi\Delta_\phi}{\epsilon},\,0\biggr\},
\end{aligned}
\end{equation}
with $T=\frac{e^t}{4\epsilon C}$, and $C$ is the Schwarzian coupling constant. $-\Delta_{\phi}/\epsilon$ is the $\bra{\phi}B\ket{\phi}$ given above. Initially, the channel capacity increases because the wormhole has not been completely opened, and it reaches its peak value at the scrambling time. After the scrambling time, the capacity decreases since the backreaction of the $\phi$ particles becomes stronger and destroys the wormhole.

We can also include finite $\alpha'$ corrections in the shockwave scattering as in Refs.~\cite{Maldacena:2017axo,Shenker:2014cwa} by changing the Dray--'t~Hooft $S$-matrix
\begin{equation}
    S=e^{i G_N \hat{P}_+\hat{P}_-}\rightarrow e^{-G_N (-i \hat{P}_+\hat{P}_-)^{1-a}},~~0\leq a\leq1,
\end{equation} this will lead to a slowdown of the growth of the quantum channel capacity. We plot the dependence of $-\partial_t\mathrm{OTOC}(t)$ on $t$  for both the maximal and nonmaximal chaotic (with $a=1/2$) settings in Fig.~\ref{fig:S_L}.\footnote{In Fig.~\ref{fig:S_L}, we plot the OTOC in the conventional normalization, i.e. normalized by $\braket{\phi\phi}$ and $\braket{OO}$. This differs from the definition in Eq.~\eqref{myOTOC} by an additional factor of $\braket{OO}$, therefore the curve starts at $1$ and then decreases. }Notice that the stringy corrections not only slow down the growth of $C_Q(\mathcal{N})$ but also decrease its maximum value over time. Detailed calculation of the stringy correction is placed in Appendix~\ref{app:stringy}.

\section{Comment on operator size interpretation}

The traversable wormhole protocol can also be understood via the concept of teleportation by size~\cite{Brown:2019hmk,Jafferis:2022crx,Kobrin:2023rzr,Tian-GangZhou:2024vxm,Nezami:2021yaq}. Consider a system with $N$ qubits, a general operator $\rho_\beta^{\frac{1}{2}}\phi$ can be expanded in Pauli basis
\begin{equation}
    \rho_\beta^{1/2}\phi=2^{-N/2}\sum_Pc_PP.
\end{equation}
Here $P$ is an $N$-qubit Pauli operator. The winding size distribution $q(\ell)$ is defined as summing the squares of these coefficients with fixed operator size $\ell$ (number of nontrivial Pauli operators in $P$). The perfect size winding requires the phase of the distribution to only depend on $|P|$ and that the phase dependence on size is linear with coefficient $\alpha$
\begin{equation}
q(\ell)=\sum_{|P|=\ell}c_P^2=\sum_{|P|=\ell}|c_P|^2e^{2i\alpha\ell/N}=\mathcal{P}(\ell)e^{2i\alpha\ell/N}.
\end{equation}
Here $\mathcal{P}(\ell)=\sum_{|P|=\ell}|c_P|^2$ is the size distribution~\cite{Zhang:2022fma}. In the wormhole setting, inserting an operator on the right side of the black hole creates a state $\phi_R|\text{TFD}\rangle$ which schematically is an operator $\rho^{1/2}_\beta\phi$. In~\cite{Maldacena:2017axo}, an observable which probes the information transfer is introduced $\mathcal{C}=\langle \text{TFD}|\phi_L\e^{i gV}\phi_R|\text{TFD}\rangle$. In a holographic teleportation setting where two SYK models with Majorana fermions $\psi^i_{L,R}$, one chooses the operator
\begin{equation}
    V(0)=-\frac{i}{N}\sum_{i=1}^N\psi^i_L\psi^i_R=1-2\frac{\ell}{N}
\end{equation} that measures the operator size.\footnote{Note that we choose the infinite temperature TFD state to be the Fermi sea vacuum of the complex fermions $c_j=\frac{1}{2}(\psi^j_L+i\psi^j_R)$.}
Then $\mathcal{C}$ is simply a Fourier transform of the winding size distribution $q_\ell(t)$ of the time evolved state $ \rho_\beta^{1/2}\phi(t)$:
\begin{equation}
    \mathcal{C}=\sum_{\ell}\e^{ig(1-\frac{2\ell}{N})}q_\ell(t).
\end{equation}

Sometimes, it was argued that perfect size winding is the feature of a traversable wormhole in Einstein gravity~\cite{Jafferis:2022crx}.\footnote{But see comments in~\cite{Kobrin:2023rzr,Jafferis:2023moh,Perugu:2025vty}.} Eq.~\eqref{eqn:QN} for the quantum channel capacity suggests an additional criterion for semiclassical traversable wormholes. In \eqref{eqn:QN}, the growth of $C_Q(\mathcal{N})$ is controlled by the time derivative of an OTOC with both $\phi$ insertions on the right. In the operator size language, this OTOC probes only the average operator size $\mathcal{P}_t(\ell)$ of the state $ \rho_\beta^{1/2}\phi(t)$. Accordingly, $C_Q(\mathcal{N})$ depends only on operator size growth and is insensitive to the winding phase
\begin{equation}
\begin{aligned}
    C_Q(\mathcal{N})&=-\beta g\partial_t \frac{\langle\phi_R| V |\phi_R\rangle}{\langle\phi_R|\phi_R\rangle}+I^0_c\\
    &=\beta g \frac{2\sum_{\ell} \mathcal{P}'_t(\ell)\ell}{N\sum_\ell \mathcal{P}(\ell)}+I^0_c.
\end{aligned}
\end{equation}
Although we are not aware of any direct bound on the growth of the wormhole signal $\mathcal{C}$, the chaos bound implies that $C_Q(\mathcal{N})$ cannot grow faster than in Einstein gravity. In this sense, the quantum channel capacity $C_Q(\mathcal{N})$, together with the associated OTOC, can serve as a quantitative benchmark for successful quantum simulations of traversable wormholes with an Einstein gravity dual.

\begin{acknowledgments}
We thank Ning Bao, Aidan Chatwin-Davies, Xi Dong, Jason Pollack, Sirui Shuai, Jinzhao Wang, Shunyu Yao, Pengfei Zhang, and Yuzhen Zhang for discussions. We thank Bryce Kobrin and David Kolchmeyer for helpful comments on the draft. Z. Y. is supported by NSFC Grants No.~12475071, No.~12447108, No.~12342501, and No.~12247103, and supported by Tsinghua University Initiative Scientific Research Program and Tsinghua University Dushi Program (No.~20251080043 and No.~20261080110).
\end{acknowledgments}

\section*{Data Availability}
There are no publicly available research data or software supporting this manuscript. Requests for further information or data should be sent to the authors.

\appendix
\section{Replica Calculation of Quantum Channel Capacity}\label{app:replica}

\begin{figure}[tbp]
    \centering
    \includegraphics[width=0.76\linewidth]{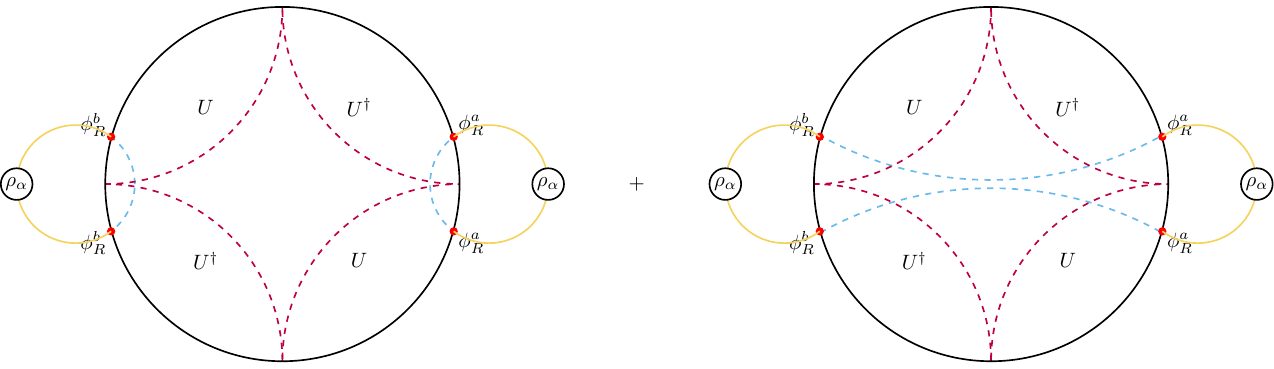}
    \caption{$n=2$ replicated geometry related to $\text{Tr}(\rho_L^2)$. The dashed purple lines represent the double trace operator, and the dashed blue lines represent the contraction between $\phi_R^a$s. The yellow lines represent the state $\rho_{\alpha}$. Here we have two different kinds of contractions of the $\phi$ particles. On the left, we have $(\text{Tr}\rho_{\alpha})^2$. In the right figure, the Wick contraction leads to $\text{Tr}\rho_{\alpha}^2$.}
    \label{fig:replica_n=2}
\end{figure}

\begin{figure}[tbp]
\centering
\begin{minipage}[t]{0.48\linewidth}
    \centering
    \includegraphics[width=0.89\linewidth]{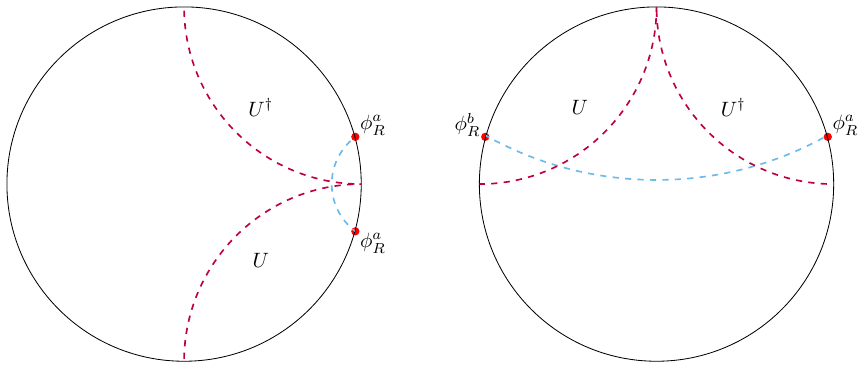}
    \caption{The left figure denotes $Z_{\text{matter}}(n)$, and the right one denotes $\tilde{Z}_{\text{matter}}(n)$.}
    \label{ZZtilde}
\end{minipage}\hfill
\begin{minipage}[t]{0.48\linewidth}
    \centering
    \includegraphics[width=0.89\linewidth]{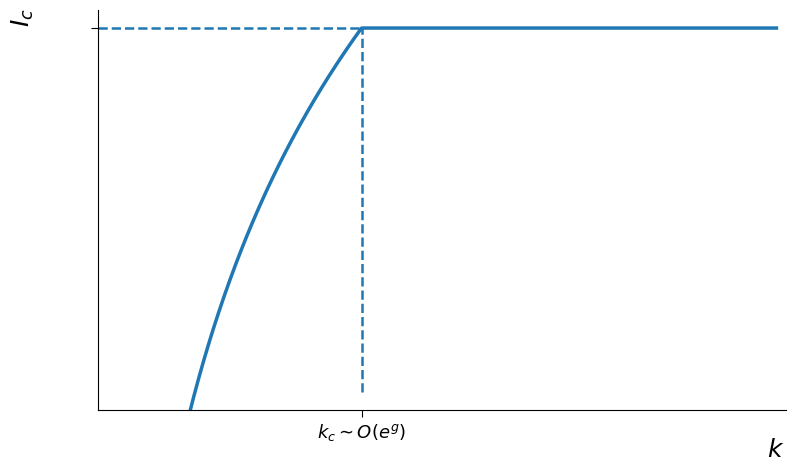}
    \caption{The coherent information versus flavor index $k$.}
    \label{coherent}
\end{minipage}
\end{figure}

In this appendix, we provide a detailed calculation for Eq.~\eqref{saturate} in JT gravity
\begin{equation}
    \max_{\text{A}} I_c(M_r\rangle L)=\beta\braket{B}.
\end{equation}
We use the replica trick to calculate the coherent information, and we show that by maximizing it over the initial states, the entropies of both sides saturate the Bekenstein bound, thus we can use the expectation value of boost operator to obtain the single-letter channel capacity. We will work in the eikonal region, which means $G_N\rightarrow 0, t\rightarrow \infty,$ with $G_N e^t$ fixed. Let us use the replica trick to calculate $S(L)$
\begin{equation}\label{eqn:SL}
    \begin{aligned}
    S(L)&=-\lim_{n\rightarrow1}\frac{1}{n-1}\log \frac{\text{Tr}(\rho_L^n)}{(\text{Tr}\rho_{L})^n}\\
        &=-\lim_{n\rightarrow1}\frac{1}{n-1}\log\left(\frac{Z(n)}{Z_1^n}\right),
    \end{aligned}
\end{equation}
where $Z(n)$ is the gravitational path integral on the $n$-replicated geometry, and $Z_1$ is the path integral on the ordinary geometry. In Fig.~\ref{fig:replica_n=2}, we show the configurations with $n=2$. Each disk has periodicity $n\beta$, and each represents a saddle in the Euclidean path integral of $Z(n=2)$.
In the eikonal region, the gravitational path-integral can be factorized into the vacuum contribution and the interaction part.
\begin{equation}
    Z(n)=Z_{o}(n)\cdot Z_{\text{int}}(n).
\end{equation}
Here $Z_{o}(n)$ denotes the partition function without operator insertions, which leads to the generalized entropy of the thermal state. It is the same for $S_L$ and $S_R$, thus will not contribute to the coherent information. $Z_{\text{int}}(n)$ denotes the contribution from insertions of $\phi$ and $U$, we take the large $K$ limit in the double trace deformation so that we can only focus on the leading contraction,

\onecolumngrid
\begin{equation}
    \braket{e^{\frac{ig}{K}\sum_{i}^{K}O_{L}^{i}O_{R}^{i}}}=\sum_n\left\langle{\frac{(ig)^n}{n!K^n}\left(\sum_{i=1}^{K}O_L^iO_R^i\right)^n}\right\rangle=\sum_n\left\langle{\frac{(ig)^n}{n!K^n}\left(\sum_{i=1}^{K}\wick{\c O_L^i \c O_R^i}\right)^n}\right\rangle=e^{ig\braket{O_LO_R}},
\end{equation}
where we ignore the subleading contractions such as $\wick{\sum_{i}O_L^{i} \c O_R^{i}\ldots\sum_j \c O_L^{j} O_R^{j}}$, which is suppressed by $1/K$ factor. And these subleading contractions correspond to the contractions among different replicated parts on a $n$-replicated geometry. Therefore, we only keep the interaction among $O$s within the same replicated part. To calculate the path-integral on the replicated geometry, we have the structure
\begin{equation}\label{Renyi}
    \text{Tr}(\rho_L^n)=\frac{Z_{o}(n)}{Z_o(1)^n\braket{\phi\phi}^n}\cdot\big(  Z_{\text{matter}}^n(n)+e^{-(n-1)S_n(\alpha) }\tilde{Z}_{\text{matter}}^n(n)+\cdots\big),
\end{equation}
where $e^{-(n-1)S_n(\alpha)}=\text{Tr}\rho_{\alpha}^n$ is the $n$th Renyi of the reference state $\rho_{\alpha}=\sum_a |\alpha^a|^2|\chi_a\rangle\langle\chi_a|$. Similarly, for the right side, we have
\begin{equation}
    \text{Tr}(\rho_R^n)=  \frac{Z_{o}(n)}{Z_o(1)^n\braket{\phi\phi}^n}\cdot\big(  \tilde{Z}_{\text{matter}}^n(n)+e^{-(n-1)S_n(\alpha)}Z_{\text{matter}}^n(n)+\cdots\big).
\end{equation}
The ellipsis represents the other contractions that are not $\mathbf{Z}_n$ symmetric and will not dominate in the calculation, but could contribute during the transition of the symmetric saddles \cite{Penington:2019kki}. The interactions factorize between different replica copies, and locally, it is described by $Z_{\text{matter}}(n)$ or $\tilde Z_{\text{matter}}(n)$. $Z_{\text{matter}}(n)$ denotes the contribution of interactions in the left half (or right half) of the first figure in Fig.~\ref{fig:replica_n=2}, and $\tilde Z_{\text{matter}}(n)$ denotes the interactions in the upper half (or lower half) of the second figure in Fig.~\ref{fig:replica_n=2}.
We show these in Fig.~\ref{ZZtilde}. Each one is calculated on the disk with periodicity $n\beta$ (or $2\pi n$ when we set $\beta=2\pi$). In order to compute $Z_{\text{matter}}(n)$, we need to compute the six-point function on the replicated geometry:
\begin{equation}\label{6pt_correlation}
    Z_{\mathrm{matter}}(n)\equiv \langle\mathcal{C}\phi(u_1)\phi(u_2)e^{igO(u_3)O(u_4)}e^{-igO(\tilde{u}_3)O(\tilde{u}_4)}\rangle_{n\beta},
\end{equation}
where $\mathcal{C}$ is the path ordering, as in Fig.~\ref{ZZtilde}. To calculate it, we use the eikonal resummation~\cite{Maldacena:2016upp,Haehl:2021tft}
\begin{equation}\label{eikonal}
Z_{\text{matter}}(n)=\frac{C}{\pi}\int dX^-dX^+ e^{-2iCX^+X^-}\langle \phi(u_1)\phi(u_2)\rangle_{+} e^{ig\langle O(u_3)O(u_4)\rangle_{-}} e^{-ig\langle O(\tilde{u}_3)O(\tilde{u}_4)\rangle_{-}}.
\end{equation}
We choose the time coordinates of these insertions to be\footnote{The $\epsilon$ is the UV regularization of $\phi$, which was chosen such that the short distance singularity is independent of $n$.}
\begin{equation}\label{timecoord}
        \begin{aligned}
        &u_1=-i\epsilon \frac{2\pi}{\beta},\ u_2=-i(-\epsilon)\frac{2\pi}{\beta},\\
        &u_3=-i(0+it)\frac{2\pi}{n\beta},\ u_4=-i(-\frac{\beta}{2}+it)\frac{2\pi}{n\beta},\\
        &\tilde{u}_3=-i(0+it)\frac{2\pi}{n\beta},\ \tilde{u}_4=-i(-\frac{\beta}{2}-(n-1)\beta+it)\frac{2\pi}{n\beta}.
    \end{aligned}
\end{equation}
We work in JT gravity, where $C$ is the Schwarzian coupling. $X^-$ is the shockwave created by $\phi$ particles and $X^+$ is created by $O$ particles. The two-point correlation functions on the fixed shockwave geometry in JT gravity are:
\begin{equation}
 \langle\phi(u_1)\phi(u_2)\rangle_{+}=\left[\frac{-i}{2\sinh\frac{u_{12}}{2}-X^+e^{-(u_1+u_2)/2}}\right]^{2\Delta_\phi},\ \
        \langle O(u_3)O(u_4)\rangle_{-}=\left[\frac{-i}{2\sinh\frac{u_{34}}{2}-X^-e^{(u_3+u_4)/2}}\right]^{2\Delta_O},
\end{equation}
where $u_{ij}=u_i-u_j$ is the Lorentz time separation. Plugging in the time coordinates, we have
\begin{equation}
\begin{aligned}
        Z_{\text{matter}}(n)= \frac{C}{\pi}\int &dX^-dX^+ e^{-2iCX^+X^-}\left[\frac{i}{2i \sin \epsilon+X^{+}}\right]^{2\Delta_{\phi}}\\
         &\exp\left(ig\left[\frac{i}{2i\sin\frac{\pi}{2n}+X^{-}e^{\frac{i\pi}{2n}+\frac{t }{n}}}\right]^{2\Delta_O}-ig\left[\frac{i}{2i\sin\frac{\pi}{2n}-X^{-}e^{\frac{-i\pi}{2n}+\frac{ t }{n}}}\right]^{2\Delta_{O}}
\right),
\end{aligned}
\end{equation}
where we have taken $\beta=2\pi$ and we employ this substitution for the rest of this appendix.
In the $n\rightarrow 1$ limit, this leads to:
\begin{equation}
     Z_{\text{matter}}(n\rightarrow 1)=\langle \phi\phi\rangle\bigg(1-(n-1)\Delta\langle  K_L\rangle\bigg),
\end{equation}
where $\Delta\langle  K_L\rangle$ is the expectation value of the vacuum-subtracted modular Hamiltonian. It can be expressed via eikonal resummation:
\begin{equation}
    \Delta\braket{K_L}=\frac{C}{\pi\braket{\phi\phi}}\int dX^{-}dX^{+}e^{-2iCX^{+}X^{-}}\left(\frac{i}{2i\sin\epsilon+X^{+}}\right)^{2\Delta_\phi}\frac{2\pi g\Delta_O X^{-}e^{t}}{(2+X^{-}e^{ t})^{2\Delta_O+1}},
\end{equation}
and can be integrated out explicitly using the confluent hypergeometric function,
\begin{equation}
    \Delta\braket{K_L}=4\pi g\Delta_O\Delta_\phi\frac{2^{2\Delta_\phi-2\Delta_O}}{T^{2\Delta_\phi}}
    U(2\Delta_\phi+1,2(\Delta_\phi-\Delta_O)+1,\frac{2}{T}),
\end{equation}
where $T=\frac{e^{t}}{4\epsilon C}$, and we set $\beta=2\pi$ for convenience.
Similarly, we have
\begin{equation}
     \tilde Z_{\text{matter}}(n\rightarrow 1)=\langle \phi\phi\rangle\bigg(1-(n-1)\Delta\langle  K_R\rangle\bigg),
\end{equation}
where $\Delta \braket{K_R}$ can be obtained in a similar way:
\begin{equation}
    \Delta\braket{K_R}=\frac{2\pi \Delta_{\phi}}{\epsilon}-4\pi g\Delta_O\Delta_\phi\frac{2^{2\Delta_\phi-2\Delta_O}}{T^{2\Delta_\phi}}
    U(2\Delta_\phi+1,2(\Delta_\phi-\Delta_O)+1,\frac{2}{T}).
\end{equation}
This leads to the generalized entropies:
\begin{equation}
    S(L)=S_0+\text{min}(\Delta\langle  K_L\rangle,S(\alpha)+\Delta\langle  K_R\rangle);~~~ S(R)=S_0+\text{min}(S(\alpha)+\Delta\langle  K_L\rangle,\Delta\langle  K_R\rangle),
\end{equation}
\twocolumngrid
where $S_0$ is the entropy that comes from the $Z_o$ part, which is the generalized entropy of the original thermal state. We have considered the case where both $\Delta\langle  K_L\rangle$ and $\Delta\langle  K_R\rangle$ are large but finite, so that only one saddle point will dominate.
For the maximization of the coherent information, we discuss two cases:
\begin{enumerate}
    \item[(i)] In the region of $\Delta\langle  K_L\rangle\gg \Delta\langle  K_R\rangle$, $S(R)$ is always given by $S_0+\Delta\langle  K_R\rangle$, so the maximization is achieved by choosing the state $\alpha$ such that $S(L)$ saturates $S_0+\Delta\langle  K_L\rangle$. This leads to Eq.~\eqref{saturate}. Now for $S(L)$, we have $0\leq S(\alpha)\leq \log k$, so if $\log k \leq \Delta\langle  K_L\rangle-\Delta\langle  K_R\rangle$, the coherent information is bounded by $\log k$. When $\log k > \Delta\langle  K_L\rangle-\Delta\langle  K_R\rangle$, which is the region mainly considered in this work, both sides saturate the Bekenstein bound, leading to Eq.~\eqref{saturate}. The relation between the coherent information and the flavor index $k$ is plotted in Fig.~\ref{coherent}. The transition point $k_c$ is given by $e^{\Delta\langle  K_L\rangle-\Delta\langle  K_R\rangle}\sim e^{O(g)}$.
    \item[(ii)] In the region of $\langle \Delta K_L\rangle\ll \langle \Delta K_R\rangle$, $S({L})$ is always given by $S_0+\Delta\langle  K_L\rangle$, and given that $S(R)\geq S_0+\Delta\langle K_L\rangle$, the coherent information is always negative, and its maximum is zero.
\end{enumerate}
Together, they lead to our final result for the quantum channel capacity Eq.~\eqref{eqn:QN}.
Notice that if we ignore the backreaction of the signal, we have
\onecolumngrid
\begin{equation}
\begin{split}
       2\pi\langle  B\rangle-I_c^0&=\frac{2 C}{\pi\braket{\phi\phi}}\int dX^{-}dX^{+}e^{-2iCX^{+}X^{-}}\left(\frac{i}{2i\sin \epsilon+X^{+}}\right)^{2\Delta_\phi}\frac{ 2\pi g\Delta_O X^{-}e^{ t}}{(2+X^{-}e^{ t})^{2\Delta_O+1}}\\
       &\approx -\frac{2\pi}{2^{2\Delta_O+1}}\frac{g\Delta_{O} e^{t}}{C}\langle P_+\rangle_{\phi}=2\pi a^+\langle P_+\rangle_{\phi},
\end{split}
\end{equation}
where we defined
\begin{equation}
    a^+=-\frac{1}{2^{2\Delta_O+1}}\frac{1}{C}g\Delta_{O} e^{t};~~~\langle P_+\rangle_{\phi}=\int \frac{dP_+\,dX^+}{2\pi}e^{\mathrm{i} P_+ X^+} \langle e^{-\mathrm{i} \hat P_+ X^+}\rangle_{\phi} P_+.
\end{equation}
Therefore, we recover Eq.~(E.97) in~\cite{Maldacena:2017axo}, that states the change of mutual information is bounded above by $2\pi a^+\langle P_+\rangle_{\phi}$ (see notation convention in~\cite{Maldacena:2017axo}, in particular here $A,B$ denote causal wedges of the left black hole after and before the double trace deformation):
\begin{equation}
    I(A,M_r)-I(B,M_r)=S(A)-S(\bar A)-(S(B)-S(\bar B))=2\pi a^+\langle P_+\rangle_{\phi}.
\end{equation}

\section{Generalization to Other Large \texorpdfstring{$N$}{N} Chaotic Systems}\label{app:scramblon}
The discussion of the relation between the channel capacity and the time derivative of OTOC can be applied to a more general class of large $N$ chaotic models whose scrambling dynamics are controlled by the scramblon effective field theory~\cite{Gu:2021xaj,Zhang:2022fma,Liu:2024nhs}. Let us consider the R\'enyi entropy calculation of Appendix~\ref{app:replica}, for example Eq.~\eqref{eqn:SL}. In the large $N$ limit, the ordinary backreaction (no large time enhancement) of matter on the thermal background can be ignored, which leads to the same factorized structure of path-integral as in Eq.~\eqref{Renyi}. The structure takes the form of a product between a thermal partition function and OTOC type correlators. For the OTOC type correlators, we can use the scramblon theory to calculate. We start with the correlator $Z_{\mathrm{matter}}(n)$ in (\ref{6pt_correlation}), which is an intermediate quantity approaching $S(L)$ and can be computed via scramblon diagrammatics. We expand the two conjugate double trace insertions and sum over all the scramblon diagrams. Propagators are drawn between each two pairs of operators which are in OTOC configuration. In our case, $O(u_3)O(u_4)$ and $O(\tilde{u}_3)O(\tilde{u}_4)$ can both form OTOC configuration with $\phi(u_1)\phi(u_2)$.

The two propagators are
\begin{equation}
\begin{aligned}
    &\lambda(t)=C^{-1}\exp\left[\frac{\varkappa}{2}(-i\pi-u_1-u_2+u_3+u_4)\right]=C^{-1}\exp\left[\frac{i\varkappa(1-n)\pi}{2n}\right]\exp\left(\frac{2\pi\varkappa}{n\beta}t\right),\\
    &\tilde\lambda(t)=C^{-1}\exp\left[\frac{\varkappa}{2}(-i\pi-u_1-u_2+\tilde u_3+\tilde u_4)\right]=C^{-1}\exp\left[\frac{i\varkappa(1-n)\pi}{2n}\right]\exp\left[\frac{2\pi\varkappa}{n\beta}\left(t+i(n-1)\frac{\beta}{2}\right)\right].
\end{aligned}
\end{equation}
\begin{figure}[tbp]
    \centering
    \begin{minipage}{0.73\linewidth}
    \centering
    \includegraphics[width=0.49\linewidth]{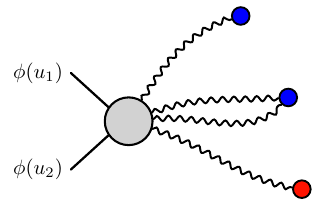}
    \caption{Scramblon diagram with $m=1$, $n=2$. The shaded blob denotes the advanced vertex and the wavy lines are the propagators. The red and blue dots denote the retarded vertices corresponding to the two OTOC configurations. The red vertex represents $igO_LO_R$ insertions from $e^{igV}$, and blue vertices represent $-ig O_L O_R$ insertions from the conjugated one.}
    \label{fig:scramblon}
    \end{minipage}
\end{figure}
Here $C$ and $\varkappa$ are model-dependent parameters. For example, in JT gravity $C\sim1/G_N$ is the Schwarzian coupling and $\varkappa=1$. In the SYK model $C\sim N$ and $\varkappa$ is related to the coupling constant in the Hamiltonian. Expanding the two exponentials by $m$ $ig$'s and $n$ $-ig$'s gives a scramblon diagram with $m$ retarded vertices $\Upsilon^{\mathrm{R},N}(u_{34})$ which should connect to the advanced vertices $\Upsilon^{\mathrm{A},N}(u_{12})$ (connected to external $\phi$ lines) via propagator $\lambda(t)$, and $n$ retarded vertices $\Upsilon^{\mathrm{R},\widetilde N}(u_{34})$ which should connect via propagator $\tilde\lambda(t)$. The superscripts $N,\,\widetilde N$ denote the number of internal propagators on that vertex. A typical diagram with $m=1$ and $n=2$ is shown in Fig.~\ref{fig:scramblon}.\par
The summation of all the diagrams is the following:
\begin{equation}\label{eq:summation_of_diagrams}
    \begin{aligned}
        Z_{\mathrm{matter}}(n)=&\sum_{m=0}^\infty\sum_{n=0}^\infty  \frac{(ig)^m}{m!}\frac{(-ig)^n}{n!}\sum_{N_1,\ldots,N_m=0}^\infty\sum_{\tilde N_1,\ldots,\tilde N_n=0}^\infty\Upsilon^{\mathrm A,\sum_pN_p+\sum_q \widetilde N_q}(u_{12})(-\lambda)^{\sum_{p}N_p}(-\tilde\lambda)^{\sum_{q}\widetilde N_q}\\
        &\times \prod_{p=1}^m\frac{1}{N_p!}\Upsilon^{\mathrm R,N_p}(u_{34})\prod_{q=1}^n\frac{1}{\widetilde N_q!}\Upsilon^{\mathrm R,\widetilde N_q}(\tilde u_{34}).
    \end{aligned}
\end{equation}
Here $N_p$ denotes the number of propagators connected between the $\phi(u_1)\phi(u_2)$ vertex and the $p$th $O(u_3)O(u_4)$ (red) vertex, and $\widetilde N_q$ has similar interpretation. For each diagram we also include a factorial to eliminate overcounted diagrams. To relate to the eikonal resummation and simplify the calculation, we use result in~\cite{Gu:2021xaj} that $\Upsilon^{\mathrm{R/A},m}(u)$ admits the following integral representation:
\begin{equation}
    \begin{aligned}
    \Upsilon^{\mathrm{R/A},m}(u)=&\int_0^\infty dy \,y^m h^{\mathrm{R/A}}(y,u)\\
        f^{\mathrm{R/A}}(x,u)=&\int_0^\infty dy\, e^{-xy} h^{\mathrm{R/A}}(y,u)=\sum_{m=0}^\infty\frac{(-x)^m}{m!}\Upsilon^{\mathrm{R/A},m}(u).
    \end{aligned}
\end{equation}
Plugging these into the expression (\ref{eq:summation_of_diagrams}), we obtain the following fully resummed form:
\begin{equation}
    \begin{aligned}
        Z_{\mathrm{matter}}(n)=&\sum_{m=0}^\infty\sum_{n=0}^\infty  \frac{(ig)^m}{m!}\frac{(-ig)^n}{n!}\int_0^\infty dy\, h^{\mathrm{A}}(y,u_{12})\\
        &\quad\times\prod_{p=1}^m \sum_{N_p}\frac{(-y\lambda)^{N_p}}{N_p!}\Upsilon^{\mathrm{R},N_p}(u_{34})\prod_{q=1}^n \sum_{\tilde N_q}\frac{(-y\tilde\lambda)^{\widetilde N_q}}{\widetilde N_q!}\Upsilon^{\mathrm{R},\widetilde N_q}(\tilde u_{34}) \\
        =&\sum_{m=0}^\infty\sum_{n=0}^\infty  \frac{(ig)^m}{m!}\frac{(-ig)^n}{n!}\int_0^\infty dy\, h^{\mathrm{A}}(y,u_{12})f^{\mathrm{R}}(\lambda y,u_{34})^mf^{\mathrm{R}}(\tilde\lambda y,\tilde u_{34})^n\\
        =&\int_0^\infty dy\, h^{\mathrm{A}}(y,u_{12})\exp\left[igf^{\mathrm{R}}(\lambda(t) y,u_{34})-igf^{\mathrm{R}}(\tilde \lambda(t) y,\tilde u_{34})\right].
    \end{aligned}
\end{equation}
We now notice that this is structurally similar to (\ref{eikonal}). As a special case, in the eikonal resummation the integration variable $y$ can be regarded as the null shift generated by the shockwave created by $\phi$-particles. Assuming $f^{\mathrm{R}}(x,u)$ is periodic in $u$ by $2\pi i$, and noting that $\lambda(t)=\tilde\lambda (t- i\beta\delta/2)$ in the replica limit $n=1+\delta$, $\delta\to 0$, we get
\begin{equation}
    Z_{\mathrm{matter}}(n)=\int_0^\infty dy\, h^{\mathrm{A}}(y,-4\pi i\epsilon/\beta)\left[1+\delta\,\frac{\beta g}{2}\,\partial_tf^R\left(C^{-1}e^{\frac{2\pi\varkappa}{\beta}t}y,-i\pi\right)\right].
\end{equation}
Completing the replica trick and repeating the same exercise for $S_R$ gives
\begin{equation}
\begin{aligned}
     C_Q(\mathcal{N})=&\max\{-\beta g \frac{\int_0^\infty dy\,h^{\mathrm{A}}(y,-4\pi i\epsilon/\beta)\partial_tf^R\left(C^{-1}e^{\frac{2\pi\varkappa}{\beta}t}y,-i\pi\right)}{\int_0^\infty dy\,h^{\mathrm{A}}(y,-4\pi i\epsilon/\beta)}-2\pi i\partial_u\log  \int_0^\infty dy\, h^{\mathrm{A}}(y,u)|_{u=-4\pi i\epsilon/\beta},0\}\\
    =&\max\{-\beta g\partial_t \mathrm{OTOC}(t)+\beta\langle\phi| B|\phi\rangle,0\}.
\end{aligned}
\end{equation}
Here we have used the scramblon field theory expression for the OTOC (\ref{myOTOC}), as derived for example in Eq.~(2.4) in~\cite{Gu:2021xaj}
\begin{equation}
    \mathrm{OTOC}(t)=\frac{\int_0^\infty dy\,h^{\mathrm{A}}(y,-4\pi i\epsilon/\beta)f^R\left(C^{-1}e^{\frac{2\pi\varkappa}{\beta}t}y,-i\pi\right)}{\int_0^\infty dy\,h^{\mathrm{A}}(y,-4\pi i\epsilon/\beta)}.
\end{equation}
Thus we have shown that the relation between the time derivative of OTOC and quantum channel capacity holds true in general chaotic theories described by the scramblon effective theory.

\section{The Calculation of Stringy Correction}\label{app:stringy}
In this appendix, we compute the stringy correction to OTOC. Working in the momentum basis, the integral form is straightforward:
\begin{equation}
\begin{aligned}
      &\langle\phi(\epsilon)O(it)\phi(-\epsilon)O(-\frac{\beta}{2}+it)\rangle\\
    =& \int dp_+dq_- e^{-\frac{1}{2C}(-ip_+q_-)^{1-a}}\langle\phi|p_+\rangle\langle p_+|\phi\rangle\langle O|q_-\rangle\langle q_-|O\rangle.
\end{aligned}
\end{equation}
The wave functions are:
\begin{equation}
    \begin{aligned}
        \langle\phi|p_+\rangle\langle p_+|\phi\rangle&=\int_{-\infty}^\infty \frac{dX^+}{2\pi}e^{i p_+X^+}\langle\phi(u_1)e^{-i\hat{P}_+X^+}\phi(u_2)\rangle,\\
        \langle O|q_-\rangle\langle q_-|O\rangle&=\int_{-\infty}^\infty \frac{dX^-}{2\pi}e^{iq_-X^-}\langle O(u_3)e^{-i\hat{Q}_-X^-}O(u_4)\rangle.
    \end{aligned}
\end{equation}
The two correlation functions under the integration are:
\begin{equation}
    \begin{aligned}
        \langle\phi(u_1)e^{-i\hat{P}_+X^+}\phi(u_2)\rangle&=\left(\frac{i}{2i\epsilon+X^+}\right)^{2\Delta_\phi}\\
        \langle O(u_3)e^{-i\hat{Q}_-X^-}O(u_4)\rangle&=\left(\frac{1}{2+X^-e^t}\right)^{2\Delta_O}.
    \end{aligned}
\end{equation}
We do the $X^{+}$ integral first, and up to some constants, we have:
\begin{equation}
    \langle\phi|p_+\rangle\langle p_+|\phi\rangle\propto (-p_+)^{2\Delta_\phi-1}e^{2\epsilon p_+}\Theta(-p_+).
\end{equation}
Thus, the final form of integral is:
\begin{equation}\label{OTOC}
\begin{aligned}
    \langle\phi(\epsilon)O(it)\phi(-\epsilon)O(-\frac{\beta}{2}+it)\rangle=\frac{1}{\Gamma(2\Delta_\phi)}&\int^0_{-\infty} dp_+\int_{-\infty}^{\infty}dq_- \int \frac{dX^-}{2\pi}e^{-\frac{1}{2C}(-ip_+q_-)^{1-a}}\\
    &(-p_+)^{2\Delta_\phi-1}e^{2\epsilon p_+}e^{iq_-X^-}\left(\frac{1}{2+X^-e^t}\right)^{2\Delta_O}.
\end{aligned}
\end{equation}
One can check that when $a\to 0$, we reproduce our previous OTOC calculation. When $a\to 1$, implying no scrambling, we obtain a simple answer that is independent of time, leading to zero entropy contribution on the left side. After including stringy corrections, the quantum Lyapunov exponent deviates from its maximal value, and the OTOC exponential decay becomes slower. Thus, we expect the channel capacity to decrease.\par
For general $a$ the integral cannot be evaluated analytically. However, for $a=1/2$ the OTOC can be obtained.\par
We first do the $q_-$ integration:
\begin{equation}
    \int_{-\infty}^{\infty}\frac{dq_-}{2\pi}e^{-\frac{1}{2C}(-ip_+q_-)^{\frac{1}{2}}}e^{iq_-X^-}=e^{\frac{p_+}{16C^2X^-}}\frac{1}{X^-}\left(\frac{ -p_+}{16\pi C^2X^-}\right)^\frac{1}{2}\Theta(X^-).
\end{equation}
Next, the $p_+$ integral gives a gamma function:
\begin{equation}
    \int^0_{-\infty} dp_+(-p_+)^{2\Delta_\phi-\frac{1}{2}}e^{\left(2\epsilon+\frac{1}{16C^2X^-}\right)p_+}=\frac{1}{(2\epsilon+\frac{1}{16C^2X^-})^{2\Delta_\phi+\frac{1}{2}}}\Gamma(2\Delta_\phi+\frac{1}{2}).
\end{equation}
So in the end, the remaining single variable integral is:
\begin{equation}
    \mathcal{A}\int_0^{\infty} \frac{dX^-}{C}\frac{1}{(X^-)^{3/2}\left(\frac{1}{16C^2X^-}+2\epsilon\right)^{2\Delta_\phi+\frac{1}{2}}(2+X^-e^t)^{2\Delta_O}}.
\end{equation}
And the constant $\mathcal{A}$ is:
\begin{equation}
    \mathcal{A}=\sqrt{\frac{1}{16\pi}}\frac{\Gamma(2\Delta_\phi+\frac{1}{2})}{\Gamma(2\Delta_\phi)}.
\end{equation}
We introduce some variables:
\begin{equation}\label{stringytime}
    x\equiv\frac{e^t}{16\epsilon C^2},\quad y\equiv X^-e^t.
\end{equation}
Then the capacity $C_Q(\mathcal{N})$ can be computed using (\ref{eqn:QN})

\begin{equation}
    \begin{aligned}
    C_Q(\mathcal{N})=\max\Biggl\{&
    \sqrt{2\pi}\,g\Delta_O\frac{\Gamma(2\Delta_\phi+\frac{1}{2})}{2^{2\Delta_O}\Gamma(2\Delta_\phi)}
    \int_0^\infty dy\,\sqrt{\frac{x}{y}}\left(1+\frac{x}{2y}\right)^{-2\Delta_\phi-\frac{1}{2}}
    \\[-2pt]
    &\hspace{4em}\times\left(1+\frac{y}{2}\right)^{-2\Delta_O-1}-\frac{2\pi\Delta_\phi}{\epsilon},\,0\Biggr\}.
    \end{aligned}
\end{equation}

The results are shown in Fig.~\ref{fig:S_L}.

\twocolumngrid

\input{draft.bbl}

\end{document}